\documentclass[runningheads]{llncs}
\usepackage{graphicx}
\usepackage{amsmath}
\usepackage{multirow}
\begin{document}
\title{Recon-GLGAN: A Global-Local context based Generative Adversarial Network for MRI Reconstruction}
\titlerunning{Recon-GLGAN: A Global-Local context based GAN for MRI Reconstruction}

% \author{Balamurali Murugesan
% \inst{1,2} \orcidID{0000-0002-3002-5845} \and
\author{Balamurali Murugesan  \thanks{Code available at  \url{https://github.com/Bala93/Recon-GLGAN}} \inst{1,2} \orcidID{0000-0002-3002-5845} \and
Vijaya Raghavan S \inst{2} 
\and
Kaushik Sarveswaran \inst{2} \orcidID{0000-0002-1525-966X}
\and 
Keerthi Ram \inst{2} \and 
Mohanasankar Sivaprakasam \inst{1,2}}
% %
\authorrunning{B. Murugesan et al.}
\institute{Indian Institute of Technology Madras (IITM), India \and
Healthcare Technology Innovation Centre (HTIC), IITM, India
\email{balamurali@htic.iitm.ac.in}}

\maketitle              % typeset the header of the contribution
\begin{abstract}
%One obvious deterrent factor accompanying these issues lies in the quality of the under-sampled k-space reconstruction. 
%The quality is hampered not only on the whole image but also on the clinically relevant regions in the image which are of more interest to the radiologists.
Magnetic resonance imaging (MRI) is one of the best medical imaging modalities as it offers excellent spatial resolution and soft-tissue contrast. But, the usage of MRI is limited by its slow acquisition time, which makes it expensive and causes patient discomfort. In order to accelerate the acquisition, multiple deep learning networks have been proposed. Recently, Generative Adversarial Networks (GANs) have shown promising results in MRI reconstruction. The drawback with the proposed GAN based methods is it does not incorporate the prior information about the end goal which could help in better reconstruction. For instance, in the case of cardiac MRI, the physician would be interested in the heart region which is of diagnostic relevance while excluding the peripheral regions. In this work, we show that incorporating prior information about a region of interest in the model would offer better performance. Thereby, we propose a novel GAN based architecture, Reconstruction Global-Local GAN (Recon-GLGAN) for MRI reconstruction. The proposed model contains a generator and a context discriminator which incorporates global and local contextual information from images. Our model offers significant performance improvement over the baseline models. Our experiments show that the concept of a context discriminator can be extended to existing GAN based reconstruction models to offer better performance. We also demonstrate that the reconstructions from the proposed method give segmentation results similar to fully sampled images. 
\end{abstract}
\section{Introduction}

Medical imaging is the preliminary step in many clinical scenarios. Magnetic resonance imaging (MRI) is one of the leading diagnostic modalities which can produce images with excellent spatial resolution and soft tissue contrast. The major advantages of MRI include its non-invasive nature and the fact that it does not use radiation for imaging. However, the major drawback of MRI is the long acquisition time, which causes discomfort to patients and hinders applications in time critical diagnoses. This relatively slow acquisition process could result in significant artefacts due to patient movement and physiological motion. The slow acquisition time of MRI can be attributed to data samples not being collected directly in the image space but rather in k-space. k-space contains spatial-frequency information that is acquired line-by-line by the MRI hardware.   In order to accelerate the MRI acquisition process, various methods ranging from Partial Fourier Imaging, Compressed Sensing and Dictionary Learning have been developed \cite{Hollingsworth_2015}. 

Recently, deep learning based methods have shown superior performance in many computer vision tasks. These methods have been successfully adapted for the MRI reconstruction problem and have shown promising results. The deep learning based methods \cite{mri_survey} for MRI reconstruction can be broadly grouped into two : 1) k-space to image domain: the fully sampled image is obtained from zero-filled k-space. Examples include AUTOMAP and ADMM-Net. 2) image to image domain: the fully sampled (FS) image  is obtained from the zero-filled (ZF) image. Our focus will be on the models of the latter kind. The work by Wang et al. \cite{wang} was the first to use convolutional neural networks to learn the mapping between ZF and FS images. %dAUTOMAP, if needed use cite mri_survey to reduce the number of references. 
%Generative Adversarial Networks (GANs) are of interest recently for tasks such as image generation and image superresolution. Introduce GAN
% A variant of a the original Generative Adversarial Network (GAN) \cite{gan} called Conditional GAN (cGAN) \cite{pix2pix} has shown promising results in many ill-posed inverse problems such as inpainting, super-resolution and denoising when compared to other deep learning based methods. The MRI reconstruction problem, having a similar problem formulation, has been approached with GANs which have yielded encouraging results. The main focus of our paper is thus the application of GANs for the MRI reconstruction problem. 
Generative Adversarial Networks (GAN) \cite{pix2pix} have shown promising results in many ill-posed inverse problems such as inpainting, super-resolution and denoising when compared to other deep learning based methods. The MRI reconstruction problem, having a similar problem formulation, has been approached with GANs and have shown encouraging results. The main focus of our paper is thus the application of GANs for the MRI reconstruction problem. 

In the GANCS work \cite{cgan}, the generator is a residual network, the discriminator is a general deep network classifier and a combination of L1 and adversarial loss constitutes the loss function. Similarly, another work ReconGAN \cite{cyclic_loss} uses a multi-stage network as a generator; a simple deep network classifier for the discriminator, and a combination of MSE loss in the image and frequency domains, adversarial loss constitute the loss function. The addition of the frequency domain loss adds data consistency. DAGAN \cite{dagan} is another work which uses U-Net as a generator, a deep learning classifier as the discriminator with a combination of MSE loss in the image and frequency domains, adversarial loss and perceptual loss as the loss function. It showed that incorporating the perceptual loss term improved the reconstructed image quality in terms of the visually more convincing anatomical or pathological details. CDFNet \cite{comgan} proposed the use of a combination of MSE loss in the image and frequency domains along with the Structural Similarity Index Measure (SSIM) as a loss function. This can be extended to a GAN setup. We will refer to this setup as ComGAN. SEGAN \cite{segan} proposed a generator network called SU-Net and used a general deep network classifier as the discriminator. The loss term used is a combination of MSE in the image domain, SSIM and patch correlation regularization.

\begin{figure}[t]
    \centering
    \includegraphics[width=0.8\linewidth]{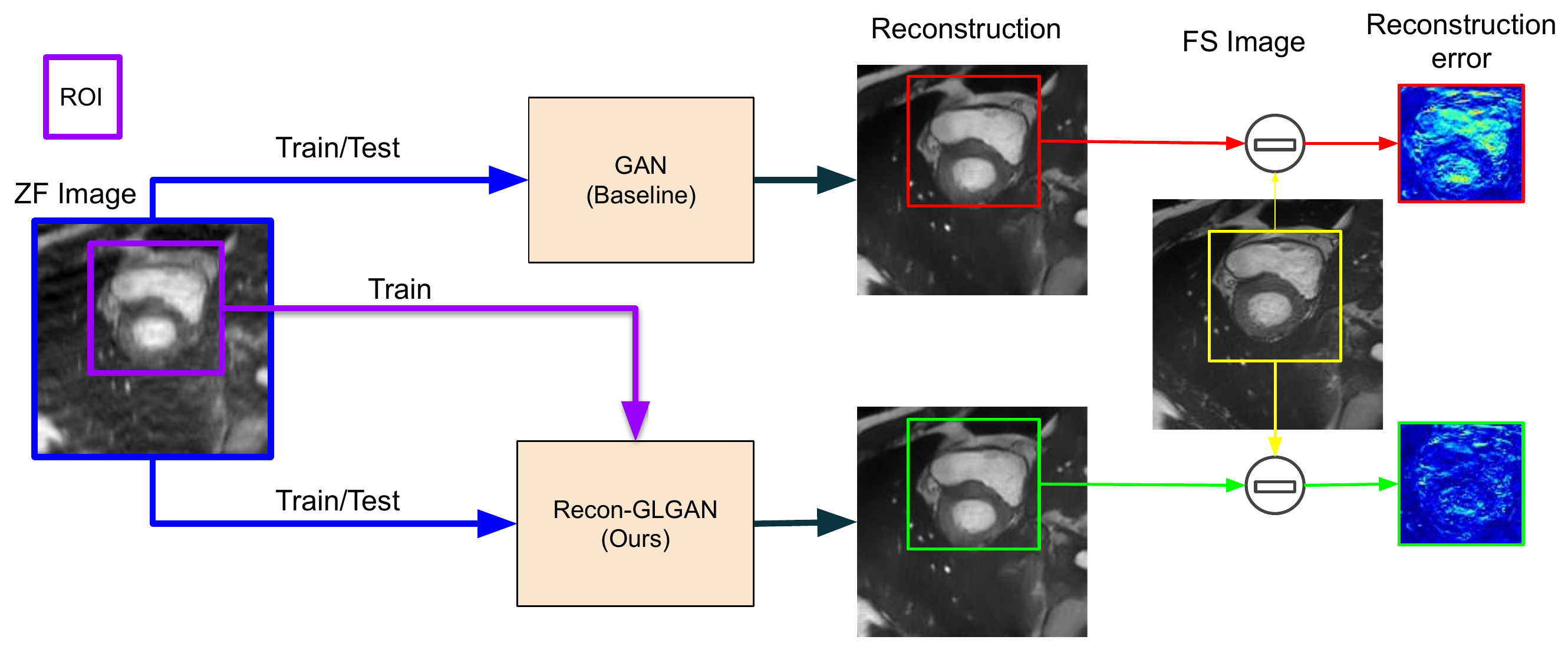}
    \caption{Illustration depicting the comparison between the baseline GAN model and our Recon-GLGAN model. In the training phase, the ZF image and the ROI are fed in as inputs to the Recon-GLGAN model, while the baseline GAN only takes the ZF image as input. In the testing stage, the ZF image is fed as input to either model to produce the reconstruction (Note: ROI is not used during testing stage). The reconstruction error of the Recon-GLGAN model is lesser than the baseline GAN model in the ROI}
    \label{fig:recon-pipeline}
\end{figure}

We refer to the concept of application-driven MRI as described in \cite{application-mri}: incorporating prior information about the end goal in the MRI reconstruction process would likely result in better performance. For instance, in the case of cardiac MRI reconstruction, the physician would be interested in the heart region, which is of diagnostic relevance while excluding the peripheral regions. Using this prior information about the region of interest (ROI) could lead to a better reconstruction. Another perspective is to note that the MRI reconstruction is not the goal in itself, but a means for further processing steps to extract relevant information such as segmentation or tissue characterisation. In general, segmentation algorithms would be interested in the specific ROI. Thus, incorporating prior information about the ROI in the reconstruction process would give two fold benefits : 
1) The reconstruction would be better, 
2) The segmentation algorithms consequently, could offer better results.
The GAN based reconstruction methods described above did not incorporate the application perspective of MRI. Recently, \cite{schlemper} proposed a method in an application-driven MRI context, where the segmentation mask is obtained directly from a ZF image. This work showed encouraging results, but the model produces only the mask as output while the physician would be interested in viewing the FS image. Incorporating the ideas stated above, we propose a novel GAN based approach for MRI reconstruction. A brief outline of our approach compared to baseline GAN approaches is shown in Figure \ref{fig:recon-pipeline}. The key contributions of our work can be summarized as follows: 
% In all the above discussed GAN based reconstruction methods, application perspective is not considered. For instance, in the case of cardiac mri reconstruction, the physician would be interested only in the heart region. The techniques discussed till now don’t use this prior information. Using this prior information can be helpful in better reconstruction. In another case, the MR reconstruction is not an end themselves, because many automated segmentation algorithms are run on the reconstructed image. The segmentation algorithm would be interested only in the specific region of interest. Using prior and giving a better reconstruction will be helpful for segmentation algorithms. Recently, a method has been proposed in application-driven MRI context, where segmentation result is obtained from an ZF image \cite{schlemper}. This work showed encouraging results, but the physician would still be interested to view the FS image which is not possible with their method.  

% Bridge this two passage with some content. 
% Contributions 
% 1. architecture + loss
% 2. extended to other GAN based networks
% 3. evaluating segmentation performance, relating to application driven MRI
% 4. extensive experimentation
\begin{enumerate}
    %\item We propose a novel GAN architecture GL-GAN with a U-Net generator and a context discriminator. The context discriminator consists of a global feature extractor, local feature extractor and a classifier. The context discriminator architecture design enables the generator to predict image considering both global and local context. We also propose a loss function which is a linear combination of L1, context adversarial loss to accommodate the proposed network. 
    %\item We conducted extensive experiments to evaluate our proposed network and idea of context discriminator. Our proposed network showed significantly better reconstruction metrics compared to the base GAN network for entire and local image region. We show that the concept of context discriminator can be easily extended to existing GAN based reconstruction architectures. The GAN architectures  with context discriminator demonstrated improved performance in all the reconstruction evaluation metrics compared to GAN with general discriminators. 
    \item We propose a novel GAN architecture, Reconstruction Global-Local GAN (Recon-GLGAN) with a U-Net generator and a context discriminator. The context discriminator consists of a global feature extractor, local feature extractor and a classifier. The context discriminator architecture leverages global as well as local contextual information from the image. We also propose a loss function which is a linear combination of context adversarial loss and L1 loss in the image domain. 
    \item We conducted extensive experiments to evaluate the proposed network with a context discriminator for acceleration factors of 2x, 4x and 8x. Our network showed significantly better reconstruction performance when compared with the baseline GAN and UNet architectures for the whole image as well as for a specific region of interest. We also show that the concept of a context discriminator can be easily extended to existing GAN based reconstruction architectures.  To this end, we replace the discriminator in the existing GAN based reconstruction architectures with our context discriminator. This showed a significant performance improvement across metrics for an acceleration factor of 4x. 
    \item We conduct preliminary experiments to show that our model produces reconstructions that result in a better performance for the segmentation task.  We demonstrate this using UNet model for segmentation, pre-trained on FS images and the corresponding masks. We observe that the segmentation results produced by the images from our Recon-GLGAN model are similar to FS images in comparison with the ZF and GAN images. 
    % This is in line with the concept of application-driven MRI described above.
    % Considering the application perspective of MRI described above, we show that our model produces reconstructions that result in better performance in a downstream task such as segmentation. We demonstrate that the Recon-GLGAN model offers better segmentation results compared to ZF and GAN when run on a popular segmentation network U-Net trained with FS images and its respective mask. 
    % with the corresponding Dice and Hausdorff metrics.
    % We show that our model produces reconstructions that result in better performance in a downstream task such as segmentation. We demonstrate this using a widely used state-of-the-art segmentation network U-Net trained with FS images and its respective mask. We observed that Recon-GLGAN model offers better segmentation results compared to ZF and GAN. 
    %that the Recon-GLGAN model offers better segmentation results compared to ZF and GAN when run on a popular segmentation network U-Net trained with FS images and its respective mask. 
\end{enumerate}
\section{Methodology}
\subsection{Problem Formulation}
% Let F denote the two-dimensional Discrete Fourier Transform matrix, also called the fully sampled Fourier encoding matrix. A $\sqrt N \times \sqrt N$ sized complex valued image $x_{f} \in \mathbb{C}^{N}$, represented in a column-wise stacked vector form is obtained by taking Inverse Discrete Fourier Transform (IDFT) of the fully sampled k-space denoted as $y_{f} \in \mathbb{C}^{N}$, since $y_{f}=F_{f}x_{f}$ . The problem now is to reconstruct $x_{f}$ from under-sampled k-space measurements, denoted by $y_{u} \in \mathbb{C}^{M}$ where $M<<N$. The problem can be represented as, $x_{u} = F_{u}^{-1}y_{u}$, where $x_{u}$ is termed as the zero-filled reconstructed image. The above equation is under-determined and makes direct inversion ill-posed. Due to sub-nyquist sampling of the k-space, the zero-filled reconstructed image suffers from aliasing. Hence motivated by the deep learning based reconstruction approaches, we formulate the problem as $\hat{x} = A(\theta)x_{u}$, where $A(\theta)$ is a deep learning network paramterized by $\theta$ that aims to reconstruct $\hat{x_{f}}$ minimizing an objective function bringing $\hat{x_{f}}$ as close as possible to the ground truth $x_{f}$.
Let $x_{f} \in {C}^{N}$ be the fully sampled complex image with dimensions $\sqrt N \times \sqrt N$ arranged in column-wise manner. $x_{f}$ is obtained from fully sampled k-space measurements ($y_{f} \in {C}^{N}$) through a fully sampled encoding matrix $F_{f}$ using the relation $y_{f}=F_{f}x_{f}$. During undersampling, a subset of kspace measurements ($y_{u} \in {C}^{M}$) say ($M<<N$) only are made. This corresponds to an undersampled image $x_{u}$ by the relation $x_{u} = F_{u}^{-1}y_{u}$. $x_{u}$ will be aliased due to sub-Nyquist sampling. Reconstructing $x_{f}$ directly from $y_{u}$ is ill-posed and direct inversion is not possible due to under-determined nature of system of equations. In our approach, we use deep learning network to learn the mapping between $x_{u}$ and $x_{f}$. The neural network thus learns to minimize the error between predicted fully sampled image ($\hat{x}_{f}$) and the ground truth ($x_{f}$).
% \subsection{Generative Adversarial Networks}
% %The networks are trained via
% % a game-theoretical approach where the generator attempts to
% % fool the discriminator into misclassifying xˆ as a real image,
% % while the discriminator constantly improves its classification
% % performance to avoid being fooled
% GAN \cite{gan} consists of a generator (G) and discriminator (D). The role of the generator is to map the latent vector $z$ to the data $x$ sampled from the distribution $p_{\text{data}}$. The role of the discriminator is to distinguish the generated ($\hat{x}$) and real ($x$) samples. The generator and discriminator competes each other, where generator tries to minimize and discriminator tries to maximize the objective function. The objective function used to learn the generator and discriminator parameters ($\theta_{G}$ and $\theta_{D}$) using joint learning is as follows:
% \begin{multline}
%     \min_{\theta_{G}} \max_{\theta_{D}} L(\theta_{D},\theta_{G}) = \mathbb{E}_{\bm{x} \sim p_{\text{data}}(\bm{x})}[\log D_{\theta_{D}}(\bm{x})] + \\ \mathbb{E}_{\bm{z} \sim p_{G}(\bm{z})}[-\log( D_{\theta_{D}}(G_{\theta_{G}}(\bm{z})))]
% \end{multline}
\subsection{Generative Adversarial Networks (GAN)}
% GAN \cite{gan} consists of a generator (G) and discriminator (D). The role of the generator is to map the latent vector $z$ to the data $x$ sampled from the distribution $p_{\text{data}}$. The role of the discriminator is to distinguish the generated ($\hat{x}$) and real ($x$) samples. 
The GAN \cite{pix2pix} consists of a generator (G) and discriminator (D). The generator (G) in GAN learns the mapping between two data distributions with the help of discriminator. In the case of MRI reconstruction, the goal of the generator is to learn the mapping between the data distribution of the ZF image ($x_{u}$) and FS image ($x_{f}$). The discriminator learns to distinguish between the generated and target reconstruction. 
%The objective function for joint learning of generator and discriminator parameters ($\theta_{G}$ and $\theta_{D}$) in GAN is represented by: 
% \begin{multline}
%  \min_{\theta_{G}} \max_{\theta_{D}} L_{GAN}(\theta_{D},\theta_{G}) = \mathbb{E}_{\bm{x_{f}} \sim p_{\text{train}}(\bm{x_{f}})}[\log D_{\theta_{D}}(\bm{x_{f}})] + \\ \mathbb{E}_{\bm{x_{u}} \sim p_{G}(\bm{x_{u}})}[-\log( D_{\theta_{D}}(G_{\theta_{G}}(\bm{x_{u}})))]
% \end{multline}

\begin{figure}[t]
    \centering
    \includegraphics[width=\linewidth]{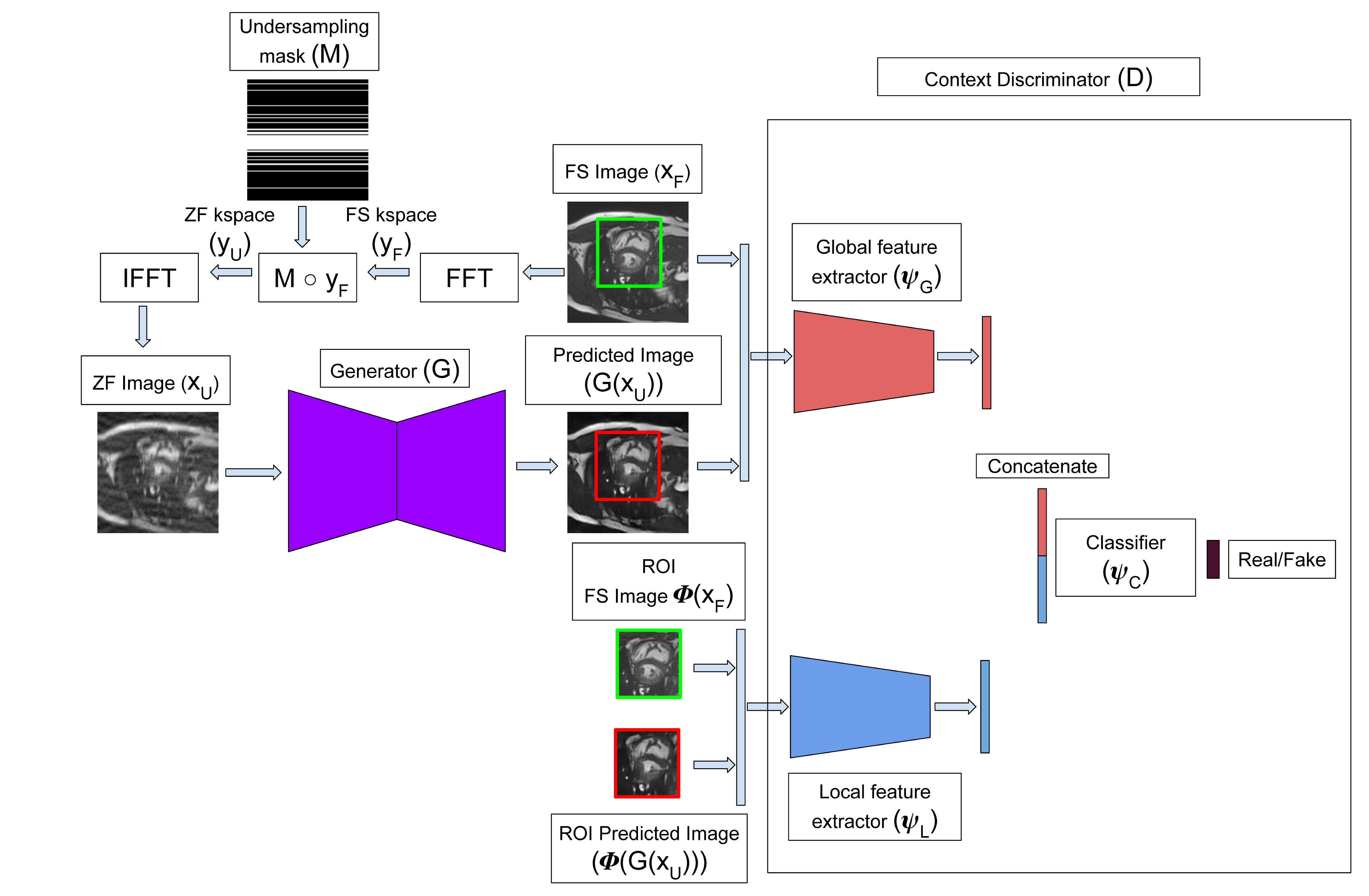}
    \caption{Recon-GLGAN architecture}
    \label{fig:arch}
\end{figure}

\subsection{Proposed Reconstruction Global-Local GAN (Recon-GLGAN)}
We propose a novel GAN architecture called Reconstruction Global-Local GAN (Recon-GLGAN). The idea is inspired from a GAN based work \cite{gl-gan} in the context of image inpainting. The idea behind Recon-GLGAN is to capture both the global and local contextual features. Recon-GLGAN consists of a generator and a context discriminator. The generator (G) tries to learn the mapping between data distribution of ZF image $x_{u}$ and FS image $x_{f}$ with the help of the context discriminator which can extract global and local features and classify it as real/fake. The context discriminator consists of three components: global feature extractor, local feature extractor and classifier. The global feature extractor ($\Psi_{G}$) takes the entire image as input while the local feature extractor ($\Psi_{L}$) takes the region of interest (ROI) ($\Phi$) from the entire image. The classifier network ($\Psi_{C}$) takes the concatenated feature vector ($\Psi_{G}(x) || \Psi_{L}(x)$) to classify the input image as real/fake. The overview of the proposed architecture is shown in Figure \ref{fig:arch}.  The joint optimization of the generator and context discriminator parameters is given by: 
%The global discriminator takes the whole image as input and classifies it to real or fake. While the local discriminator takes the cropped image and classifies it to real or fake. The objective function to optimize the parameters of generator, global and local discriminator is given by:
% \begin{equation}
%      \min_{\theta_{G}} \max_{\theta_{DG},\theta_{DL}} L_{proposed}(\theta_{DG},\theta_{DL},\theta_{G}) = L_{Global}(\theta_{DG},\theta_{G}) +  L_{Local}(\theta_{DL},\theta_{G}) 
% \end{equation}
% where 
% \begin{multline}
%  L_{Global}(\theta_{DG},\theta_{G}) = \mathbb{E}_{\bm{x_{t}} \sim p_{\text{train}}(\bm{x_{t}})}[\log D_{\theta_{DG}}(\bm{x_{t}})] + \\  \mathbb{E}_{\bm{x_{u}} \sim p_{G}(\bm{x_{u}})}[-\log( D_{\theta_{DG}}(G_{\theta_{G}}(\bm{x_{u}})))]   
% \end{multline}
% \begin{multline}
%  L_{Local}(\theta_{DL},\theta_{G}) = \mathbb{E}_{\bm{x_{t}} \sim p_{\text{train}}(\bm{x_{t}})}[\log D_{\theta_{DL}}(\bm{\phi_{crop}(x_{t})})] + \\ \mathbb{E}_{\bm{x_{u}} \sim p_{G}(\bm{x_{u}})}[-\log( D_{\theta_{DL}}(\phi_{crop}(G_{\theta_{G}}(\bm{x_{u}}))))]
% \end{multline}
\begin{multline}
 \min_{\theta_{G}} \max_{\theta_{D}} L_{Recon-GLGAN}(\theta_{D},\theta_{G}) = E_{x_{f} \sim p_{\text{train}}(x_{f})}[\log D_{\theta_{D}}(x_{f})] + \\ E_{x_{u} \sim p_{G}(x_{u})}[-\log( D_{\theta_{D}}(G_{\theta_{G}}(x_{u})))]
\end{multline}
\begin{equation}
    D_{\theta_{D}}(x) = \Psi_{C}(\Psi_{G}(x) || \Psi_{L}(\Phi(x)))
\end{equation}

\subsection{Network architecture}
% The network architecture GL-GAN is inspired from a GAN based work \cite{gl-gan} on inpainting task. The GL-GAN consists of four major components: generator , global feature extractor, local feature extractor and a classifier. The generator network takes the zero-filled image as input and predicts the corresponding fully sampled image. The global and local discriminator helps the generator to predict fully sampled image close to its target. The global discriminator takes the whole predicted/fully-sampled image as input whereas the local discriminator takes the cropped region of interest of predicted/fully-sampled image as input. Both the discriminators output a feature vector based on the input image. The extracted feature vectors are stacked and given to an Multi Layer Perceptron (MLP) for classification. The MLP classifies the input given to the discriminators as real or fake. The global discriminator helps to get the global context of the predicted/target image while the local discriminator helps to get the local context of the predicted/target image.
% The network architecture GL-cGAN is inspired from a GAN based work \cite{gl-gan} on inpainting task. The GL-cGAN consists of two major components: a generator and a context discriminator. The context discriminator consists of three components: global feature extractor, local feature extractor and classifier. 
\subsubsection{Generator (G):}
The most commonly used encoder-decoder architecture U-Net \cite{unet} is used as the generator. 
%The architecture consists of a contracting path to capture context and a symmetric expanding path that enables precise localization. The contracting path is a convolutional neural network with repeated application of two 3$\times$3 convolutions followed by a batch normalization layer, rectifier linear unit (ReLU) activation and 2$\times$2 max pooling with a stride of 2 for downsampling. The number of channels for each convolution after the max-pooling layers is increased by a factor of 2. In the expanding path, the feature maps are upsampled by a factor of 2 followed by a convolutional layer to halve the number of channels. The feature maps from the contracting path are concatenated with the corresponding convolutional layer in the expanding path. The 64-channel output feature vector from the expanding path is converted to a single channel with 1$\times$1 convolution.  
\subsubsection{Context Discriminator (D) :} \label{sect:discriminator}
 \begin{itemize}
 \item {\textit{Global feature extractor} ($\Psi_{G}$): The global feature extractor operates on the whole image. In our case, the input image dimension is 160$\times$160. The stack of 3 convolutional layers followed by 2 fully connected layers is used as the global feature extractor. Leaky ReLu is used as an activation function for each layer. Average pooling is applied after each convolutional layer. The kernel size of convolutional layer is represented by : (Output channels, Input channels, height, width, stride, padding). The three convolution layers have the following parameters: 1) (32,1,9,9,1,0) 2) (64,32,5,5,1,0) 3) (64,64,5,5,1,0). The 2 fully connected layers converts the feature maps from convolutional layer to 64-dimensional feature vector.}
\item {\textit{Local feature extractor} ($\Psi_{L}$): The local feature extractor operates on the specific ROI of an image. In our case, the dimension of the ROI is 60$\times$60. The architecture is largely similar to that of the global feature extractor except for the dimensions of the feature vector of the fully connected layer, which is modified according to the image dimensions. The output is a 64-dimensional feature vector.}
\item {\textit{Classifier} ($\Psi_{C}$)}:
The outputs of the global and the local feature extractors are concatenated together into a single 128-dimensional vector, which is then passed to a single fully-connected layer, to output a single, continuous value. A sigmoid activation function is used so that this value is in the [0, 1] range and represents the probability that the reconstruction is real/fake.
\end{itemize}
\subsection{Loss function}
The loss function to accommodate our network design is given below: 
\begin{equation}
    L_{total} = \lambda_{1} L_{imag} + \lambda_{2} L_{context}  
\end{equation}
\begin{equation} 
    L_{imag}  = E_{x_{u},x_{f}}[||x_{f}-G(x_{u})||_{1}]
\end{equation}
\begin{equation} 
    L_{context} = E_{x_{f}}[log(D(x_{f}))] + E_{x_{u}}[-log(D(G(x_{u})))]
\end{equation}
where $L_{imag}$ is the L1 loss between predicted and target fully sampled image, $L_{context}$ is the context adversarial loss. 
\section{Experiments and Results}
\subsection{Dataset}
% Our approach leverages information about the ROI to aid the reconstruction process. Thus, the dataset we use should allow us to: 
% 1) Compute the ROI, which is a bounding box obtained from the position of the segmentation mask 2) Verify that the reconstructions produced by our model result in better performance for a downstream task such as segmentation. 
Automated Cardiac Diagnosis Challenge (ACDC) \cite{acdc_dataset} is a cardiac MRI segmentation dataset. The dataset has 150 and 50 patient records for training and testing respectively. From the patient records, 2D slice images are extracted and cropped to 160$\times$160. The extracted 2D slices amount to 1841 for training and 1076 for testing. The slices are normalized to the range (0-1). In the context of MRI reconstruction, the slice images are considered as FS images while the ZF images are obtained through cartesian undersampling masks corresponding to 2x, 4x and 8x accelerations.    

The MR images in training set have their corresponding segmentation masks whereas the segmentation masks for MR images in test set are not publicly available. The dimensions of the ROI is set to 60$\times$60 based on a study of the sizes of the segmentation masks in the training set. In the training phase, the center of the ROI for each slice is the midpoint of the closest bounding box of the corresponding segmentation mask. 
% magnitude alone is considered. segmentation and edge map. 
\subsection{Evaluation Metrics}
Peak Signal-to-Noise Ratio (PSNR), Structural Similarity Index (SSIM), Normalised Mean Square Error (NMSE) metrics are used to evaluate the reconstruction quality for the entire image and its ROI. The segmentation quality is evaluated using Dice similarity coefficient (DICE) and Hausdorff distance (HD). 
\subsection{Implementation Details} 
% Pytorch, No of epochs, Learning rate batch size, SGD for discriminator, Adam for generator. Report all the training and implementation setting. GPU and system config. Add lambda values in loss functions.
The models were implemented in PyTorch. All  models were trained for 150 epochs on two Nvidia GTX-1070 GPUs. Adam optimizer was used for the generator, with a learning rate of 0.001. Stochastic Gradient Descent optimizer was used for the discriminator, with a learning rate of $5e^{-3}$. For the loss term, $\lambda_{2}= 4e^{-4}$, and $\lambda_{1}=1$. 
% \footnote{Code will be made publicly available}

The ROI for the MR images in the test set is obtained by following the algorithm described in \cite{crop_algorithm}. This ROI information is not used for inference, it is used only to evaluate the ROI's reconstruction quality.

\subsection{Results and Discussion}
\subsubsection{Reconstruction}
To evaluate the proposed network, we perform the following experiments: 

1) We compare our proposed Recon-GLGAN with the baseline architecture GAN, U-Net, and the ZF images. The metrics for each model for the whole image as well as ROI are shown in Table \ref{tab:results1}. The results show that our model Recon-GLGAN performs better than the baseline GAN and U-Net across all metrics for all acceleration factors. 
\begin{table}
\centering
\caption{Comparison of Recon-GLGAN with baseline architectures for 2x, 4x and 8x accelerations(FI-Full image)}
\label{tab:results1}
\begin{tabular}{|l|l|l|l|l|l|}
\hline
\multicolumn{3}{|l|}{} & \multicolumn{1}{c|}{NMSE} & \multicolumn{1}{c|}{PSNR} & \multicolumn{1}{c|}{SSIM} \\ \hline
\multirow{8}{*}{2x} & \multirow{4}{*}{FI} & Zero-filled & 0.01997 $\pm$ 0.01 & 26.59 $\pm$ 3.19 & 0.8332 $\pm$ 0.06 \\ \cline{3-6} 
 &  & UNet & 0.00959 $\pm$ 0.00 & 29.7 $\pm$ 2.97 & 0.9069 $\pm$ 0.03 \\ \cline{3-6} 
 &  & GAN & 0.00958 $\pm$ 0.01 & 29.72 $\pm$ 3.03 & 0.9083 $\pm$ 0.03 \\ \cline{3-6} 
 &  & Recon-GLGAN & \textbf{0.00956 $\pm$ 0.00} & \textbf{29.74 $\pm$ 3.0} & \textbf{0.9108 $\pm$ 0.03} \\ \cline{2-6} 
 & \multirow{4}{*}{ROI} & Zero-filled & 0.01949 $\pm$ 0.02 & 25.48 $\pm$ 3.73 & 0.859 $\pm$ 0.05 \\ \cline{3-6} 
 &  & UNet & 0.00952 $\pm$ 0.01 & 28.48 $\pm$ 3.03 & 0.9036 $\pm$ 0.04 \\ \cline{3-6} 
 &  & GAN & \textbf{0.00942 $\pm$ 0.00} & 28.53 $\pm$ 3.12 & 0.904 $\pm$ 0.04 \\ \cline{3-6} 
 &  & Recon-GLGAN & 0.00944 $\pm$ 0.01 & \textbf{28.54 $\pm$ 3.19} & \textbf{0.9065 $\pm$ 0.04} \\ \hline
\multirow{8}{*}{4x} & \multirow{4}{*}{FI} & Zero-filled & 0.03989 $\pm$ 0.03 & 23.65 $\pm$ 3.38 & 0.7327 $\pm$ 0.08 \\ \cline{3-6} 
 &  & UNet & 0.01962 $\pm$ 0.01 & 26.62 $\pm$ 3.209 & 0.8419 $\pm$ 0.05 \\ \cline{3-6} 
 &  & GAN & 0.01934 $\pm$ 0.01 & 26.68 $\pm$ 3.08 & 0.8465 $\pm$ 0.05 \\ \cline{3-6} 
 &  & Recon-GLGAN & \textbf{0.01905 $\pm$ 0.01} & \textbf{26.8 $\pm$ 3.25} & \textbf{0.8497 $\pm$ 0.05} \\ \cline{2-6} 
 & \multirow{4}{*}{ROI} & Zero-filled & 0.03886 $\pm$ 0.04 & 22.63 $\pm$ 3.87 & 0.7514 $\pm$ 0.07 \\ \cline{3-6} 
 &  & UNet & 0.01931 $\pm$ 0.01 & 25.46 $\pm$ 3.35 & 0.8242 $\pm$ 0.06 \\ \cline{3-6} 
 &  & GAN & 0.01925 $\pm$ 0.02 & 25.52 $\pm$ 3.38 & 0.8301 $\pm$ 0.06 \\ \cline{3-6} 
 &  & Recon-GLGAN & \textbf{0.01878 $\pm$ 0.02} & \textbf{25.66 $\pm$ 3.26} & \textbf{0.8327 $\pm$ 0.06} \\ \hline
\multirow{8}{*}{8x} & \multirow{4}{*}{FI} & Zero-filled & 0.08296 $\pm$ 0.06 & 20.46 $\pm$ 3.24 & 0.6443 $\pm$ 0.09 \\ \cline{3-6} 
 &  & UNet & 0.03353 $\pm$ 0.02  & 24.26 $\pm$ 2.71 & 0.7547 $\pm$ 0.07 \\ \cline{3-6} 
 &  & GAN & 0.03359 $\pm$ 0.02 & 24.25 $\pm$ 2.71 & 0.7557 $\pm$ 0.07 \\ \cline{3-6} 
 &  & Recon-GLGAN & \textbf{0.03286 $\pm$ 0.02} & \textbf{24.32 $\pm$ 2.68} & \textbf{0.7562 $\pm$ 0.07} \\ \cline{2-6} 
 & \multirow{4}{*}{ROI} & Zero-filled & 0.07943 $\pm$ 0.08 & 19.47 $\pm$ 3.82 & 0.6435 $\pm$ 0.07 \\ \cline{3-6} 
 &  & UNet & 0.03147 $\pm$ 0.02  &  23.31 $\pm$ 2.88 & 0.72 $\pm$ 0.07 \\ \cline{3-6} 
 &  & GAN & 0.03129 $\pm$ 0.02 & 23.33 $\pm$ 2.92 & \textbf{0.7294 $\pm$ 0.07} \\ \cline{3-6} 
 &  & Recon-GLGAN & \textbf{0.03102 $\pm$ 0.02} & \textbf{23.34 $\pm$ 2.82} & 0.7293 $\pm$ 0.07 \\ \hline
\end{tabular}
\end{table}
\begin{figure}
    \centering
    \includegraphics[width=\linewidth]{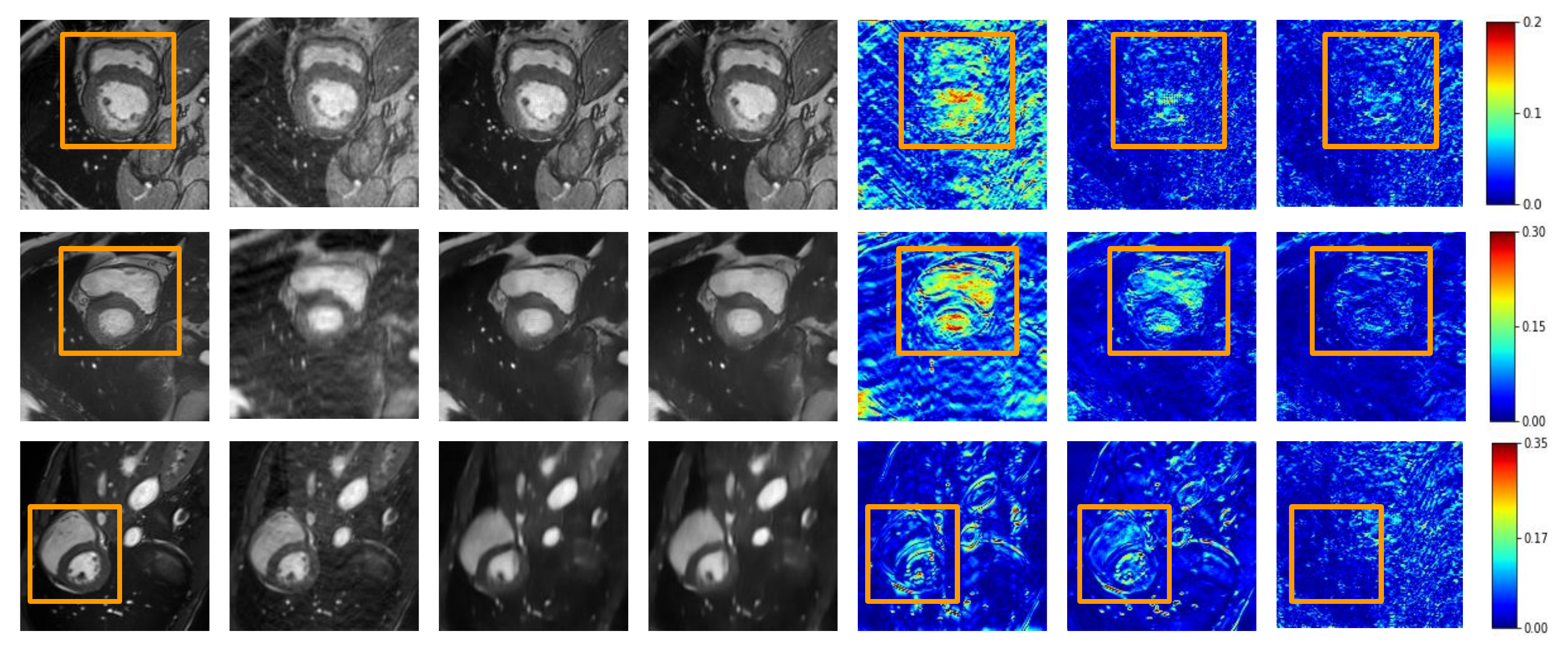}
    \caption{From Left to Right: Ground Truth FS image, ZF image, GAN reconstructed image, Recon-GLGAN reconstructed image, ZF reconstruction error, GAN reconstruction error and Recon-GLGAN reconstruction error. From Top to Bottom: Images corresponding to different acceleration factors: 2x, 4x and 8x.}
    \label{fig:cmp recon cGAN and GL-cGAN}
\end{figure}
We also note that our model offers appreciable performance improvement for 4x and 8x acceleration factors compared to 2x. This can be attributed to the fact that the image degradation in the case of 2x is not severe when compared with 4x and 8x. The qualitative comparison of ZF, GAN and Recon-GLGAN for different acceleration factors are shown in Figure \ref{fig:cmp recon cGAN and GL-cGAN}. In the Figure, it can be observed that reconstruction error of Recon-GLGAN for entire image and its ROI is better than GAN. But, it is evident that, the reconstruction error of Recon-GLGAN is significantly better than GAN in the ROI compared with the entire image. This behaviour can be attributed to the design of context discriminator which has a separate feature extraction path for specified ROI. The design of context discriminator enables the generator to specifically learn the ROI along with the entire image during the training phase. 
% In the Figure, comparing the reconstruction error in ROI of ZF, GAN and Recon-GLGAN for all % 
% ROI based MR image - helpful for this cases. 
\begin{table}[t]
\centering
\caption{GAN based reconstruction architectures and their loss terms}
\label{tab:ganarch}
\begin{tabular}{|c|c|l|}
\hline
\multicolumn{2}{|c|}{Architecture}      & \multicolumn{1}{c|}{Loss function terms}          \\ \hline
\multirow{2}{*}{ReconGAN} & -           & $L_{imag}$, $L_{global}$, $L_{freq}$              \\ \cline{2-3} 
                          & GL-ReconGAN & $L_{imag}$, $L_{context}$, $L_{freq}$             \\ \hline
\multirow{2}{*}{DAGAN}    & -           & $L_{imag}$, $L_{global}$, $L_{freq}$, $L_{vgg}$   \\ \cline{2-3} 
                          & GL-DAGAN    & $L_{imag}$, $L_{context}$, $L_{freq}$, $L_{vgg}$  \\ \hline
\multirow{2}{*}{SEGAN}    & -           & $L_{imag}$, $L_{global}$, $L_{ssim}$              \\ \cline{2-3} 
                          & GL-SEGAN    & $L_{imag}$, $L_{context}$, $L_{ssim}$             \\ \hline
\multirow{2}{*}{COMGAN}   & -           & $L_{imag}$, $L_{freq}$, $L_{global}$, $L_{ssim}$  \\ \cline{2-3} 
                          & GL-COMGAN   & $L_{imag}$, $L_{freq}$, $L_{context}$, $L_{ssim}$ \\ \hline
\end{tabular}
\end{table}

\begin{table}
\centering
\caption{Reconstruction metric comparison for full image and region of interest for various GAN based reconstruction architecture for 4x accelerations(FI - Full Image)}
\label{tab:results2}
\begin{tabular}{|c|c|c|c|c|c|}
\hline
\multicolumn{3}{|l|}{} & NMSE & PSNR & SSIM \\ \hline
\multirow{4}{*}{ReconGAN} & \multirow{2}{*}{FI} & - & 0.01857 $\pm$ 0.01 & 26.82 $\pm$ 2.89 & 0.8485 $\pm$ 0.05 \\ \cline{3-6} 
 &  & GL-ReconGAN & \textbf{0.01844 $\pm$ 0.01} & \textbf{26.91 $\pm$ 3.12} & \textbf{0.8498 $\pm$ 0.05} \\ \cline{2-6} 
 & \multirow{2}{*}{ROI} & - & \textbf{0.018 $\pm$ 0.01} & \textbf{25.76 $\pm$ 3.06} & 0.832 $\pm$ 0.06 \\ \cline{3-6} 
 &  & GL-ReconGAN & 0.01836 $\pm$ 0.01 & 25.72 $\pm$ 3.24 & \textbf{0.8336 $\pm$ 0.06} \\ \hline
\multirow{4}{*}{SEGAN} & \multirow{2}{*}{FI} & - & 0.01862 $\pm$ 0.01 & 26.84 $\pm$ 3.10 & 0.8483 $\pm$ 0.06 \\ \cline{3-6} 
 &  & GL-SEGAN & \textbf{0.01817 $\pm$ 0.01} & \textbf{27.02 $\pm$ 3.4} & \textbf{0.8545 $\pm$ 0.05} \\ \cline{2-6} 
 & \multirow{2}{*}{ROI} & - & 0.0185 $\pm$ 0.01 & 25.64 $\pm$ 3.19 & 0.8308 $\pm$ 0.07 \\ \cline{3-6} 
 &  & GL-SEGAN & \textbf{0.01793 $\pm$ 0.01} & \textbf{25.87 $\pm$ 3.56} & \textbf{0.838 $\pm$ 0.06} \\ \hline
\multirow{4}{*}{ComGAN} & \multirow{2}{*}{FI} & - & 0.01899 $\pm$ 0.01 & 26.78 $\pm$ 3.14 & 0.8481 $\pm$ 0.05 \\ \cline{3-6} 
 &  & GL-ComGAN & \textbf{0.01789 $\pm$ 0.01} & \textbf{27.06 $\pm$ 3.26} & \textbf{0.8505 $\pm$ 0.05} \\ \cline{2-6} 
 & \multirow{2}{*}{ROI} & - & 0.01872 $\pm$ 0.01 & 25.64 $\pm$ 3.28 & 0.8315 $\pm$ 0.06 \\ \cline{3-6} 
 &  & GL-ComGAN & \textbf{0.01766 $\pm$ 0.02} & \textbf{25.91 $\pm$ 3.25} & \textbf{0.834 $\pm$ 0.06} \\ \hline
\multirow{4}{*}{DAGAN} & \multirow{2}{*}{FI} & - & 0.01903 $\pm$ 0.01 & 26.75 $\pm$ 3.06 & 0.8452 $\pm$ 0.06 \\ \cline{3-6} 
 &  & GL-DAGAN & \textbf{0.01851 $\pm$ 0.01} & \textbf{26.87 $\pm$ 3.03} & \textbf{0.845 $\pm$ 0.06} \\ \cline{2-6} 
 & \multirow{2}{*}{ROI} & - & \textbf{0.01838 $\pm$ 0.01} & \textbf{25.68 $\pm$ 3.04} & 0.8272 $\pm$ 0.07 \\ \cline{3-6} 
 &  & GL-DAGAN & 0.01858 $\pm$ 0.01 & 25.62 $\pm$ 3.016 & \textbf{0.8277 $\pm$ 0.07} \\ \hline
\end{tabular}
\end{table}

\begin{figure}[t]
    \centering
    \includegraphics[width=0.8\linewidth]{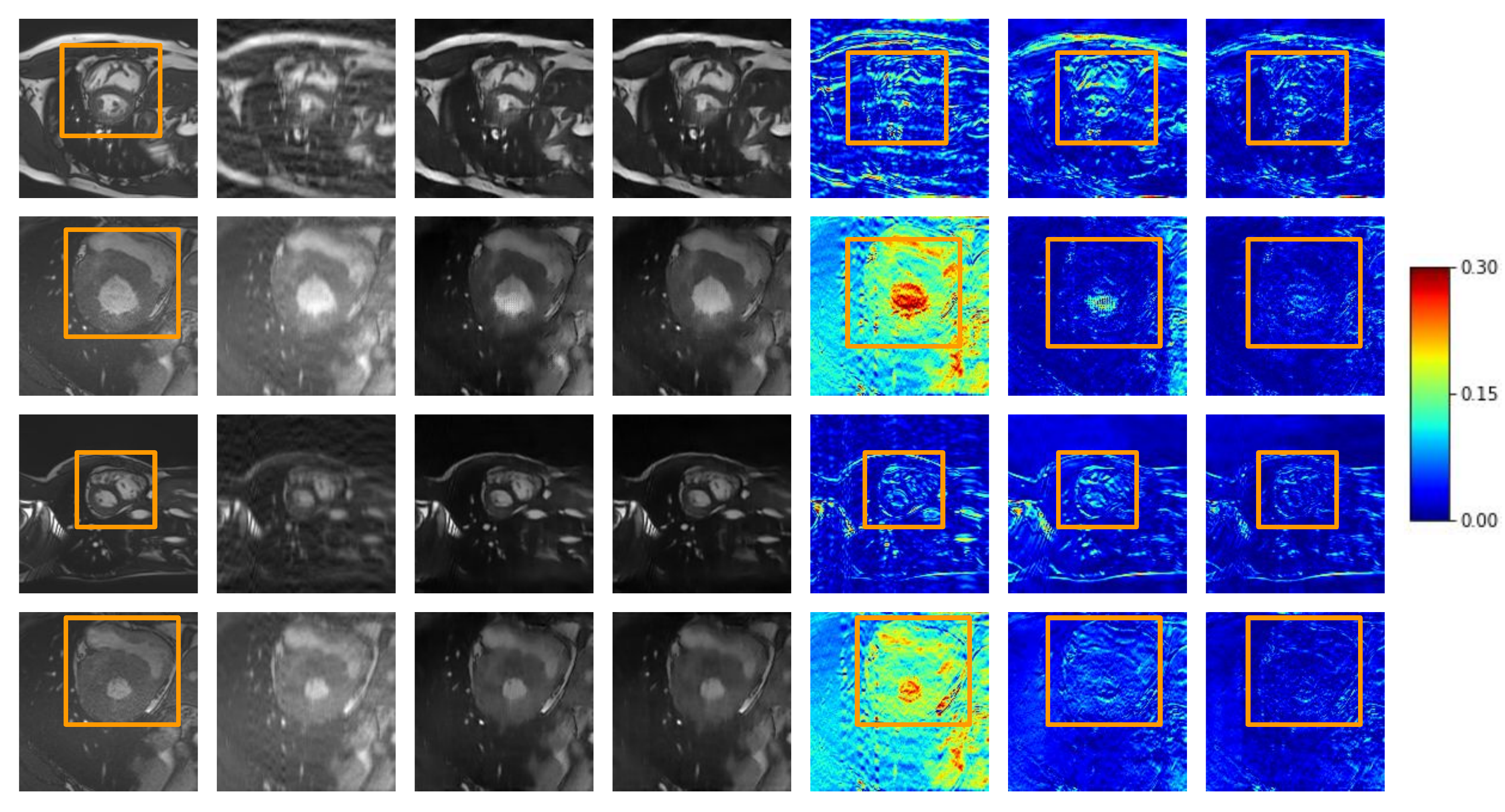}
    \caption{From Left to Right: Ground Truth FS image, ZF image for 4x undersampling factor, GAN with basic discriminator reconstructed image, GAN with context discriminator reconstructed image, ZF reconstruction error, GAN with basic discriminator reconstruction error and GAN with context discriminator reconstruction error. From top to bottom: ReconGAN, SEGAN, ComGAN, DAGAN.}
    \label{fig:cmp GL with GANs}
\end{figure}

2) We attempt to show that the concept of a context discriminator can be extended to existing GAN based works for MRI reconstruction. The different GAN based architectures and their corresponding loss functions can be found in Table \ref{tab:ganarch}. In this experiment to ensure a fair comparison, the generator is set to U-Net, discriminator is set to global feature extractor($\Psi_{G}$) followed by a classifier($\Psi_{C}$)(basic discriminator) and the loss functions are taken from their respective works \cite{dagan,comgan,segan,cgan}. This arrangement means that the difference between the various GAN based architectures comes only from the generator loss. In this experiment, we replace the basic discriminator of the GAN architectures with our proposed context discriminator. The results comparing the GAN architectures with basic discriminator and context discriminator are reported in Table \ref{tab:results2}. From the Table, it is clear that the GAN with context discriminator have shown improved results compared to GAN with basic discriminator for different generator loss. A few sample results comparing the GAN based reconstruction methods with basic and context discriminator are shown in Figure \ref{fig:cmp GL with GANs}. From the figure we observe that the ROI's reconstruction error for GAN with context discriminator is lesser compared to GAN with the basic discriminator. This shows that the context discriminator can be extended to other GAN based reconstruction methods.

% For the experiments, the models mentioned in the table use U-Net as the generator and the global feature extraction architecture described in Section \ref{sect:discriminator} is used as the discriminator, to ensure a fair comparison. % The loss functions used are adapted directly from the corresponding works \cite{dagan,comgan,segan,cgan}. We perform the following procedure as an experiment: replace the discriminator in each architecture with the proposed context discriminator and the results obtained are reported in Table \ref{tab:results2}. 

% From the Table, it is clear that there is significant performance improvement for the image as well as the ROI in most cases. Some sample results comparing the GAN based reconstruction methods with global and context discriminator are shown in Figure \ref{fig:cmp GL with GANs}. In the Figure, it can be observed that the reconstruction error of image and its ROI is lesser for methods trained with context discriminator compared to methods trained using global discriminator. The ability of the context discriminator  

\subsubsection{Segmentation} 
\begin{figure}
\centering
\includegraphics[width=0.4\linewidth]{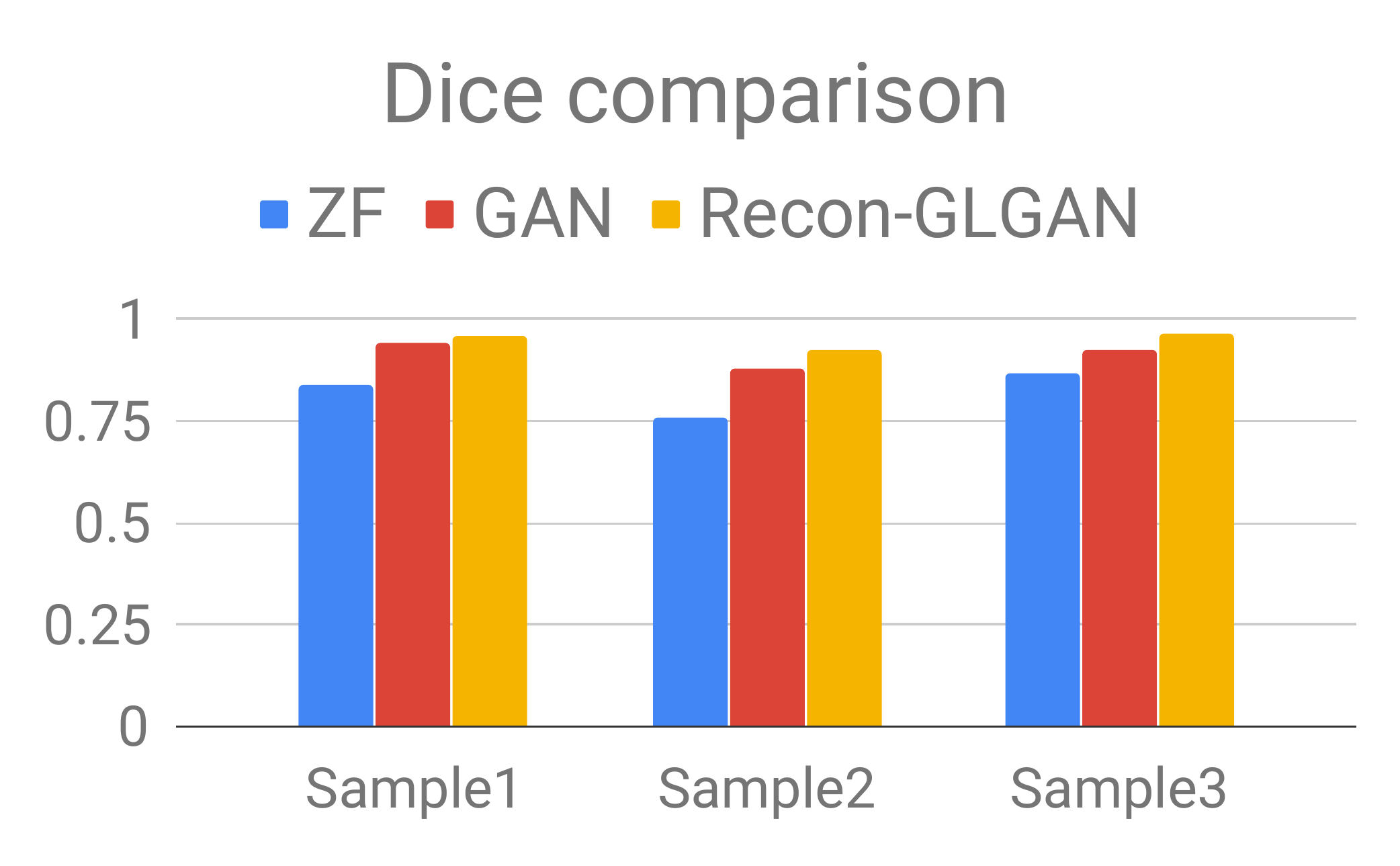}
\includegraphics[width=0.4\linewidth]{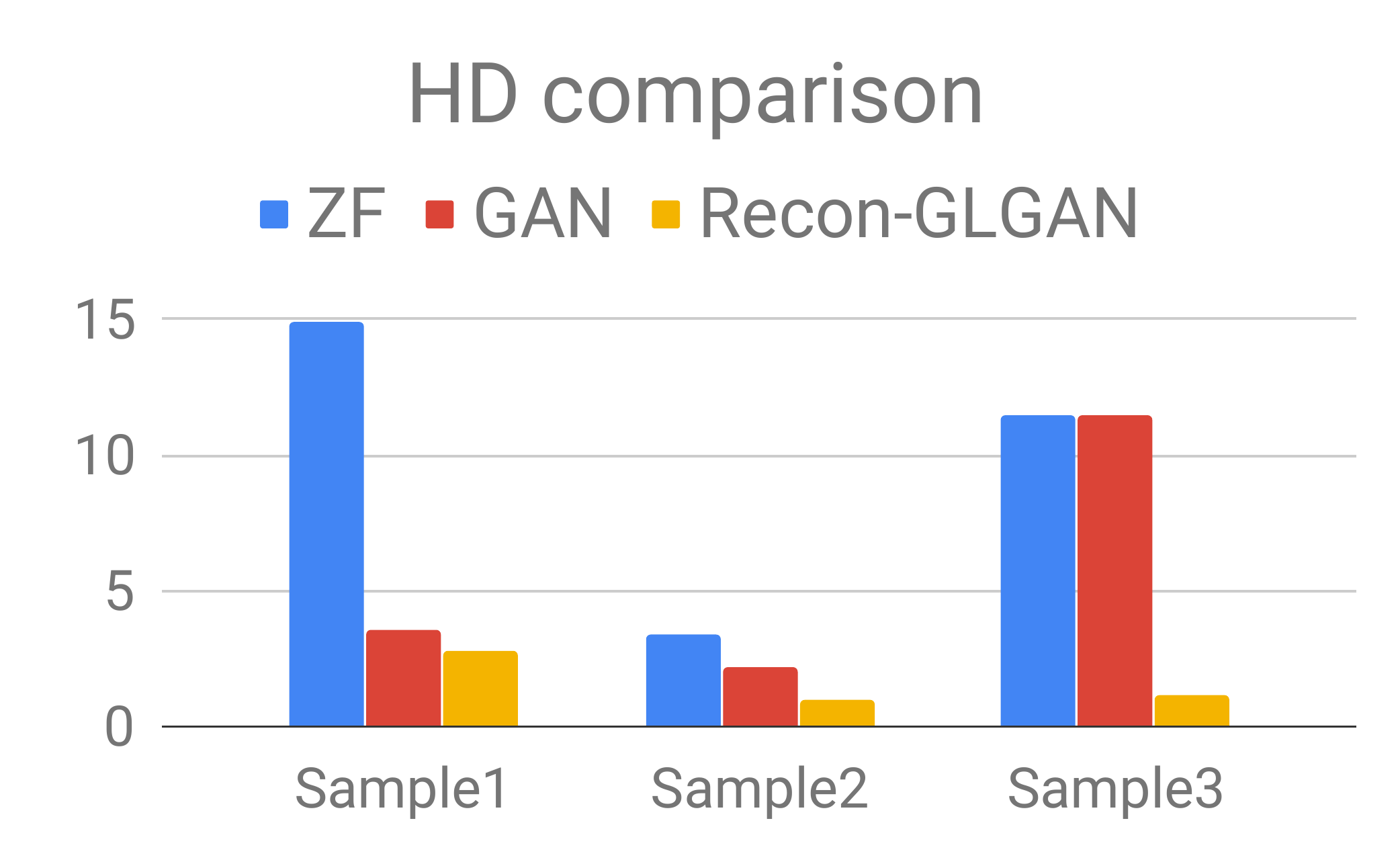}
\caption{Segmentation metrics: Dice and HD comparison for image samples 1, 2 and 3}
\label{fig:cmp seg cGAN and GL-cGAN quatitative}
\end{figure}

Image segmentation is an important task in medical imaging and diagnosis. For instance, in the case of cardiac MRI, the segmentation of left ventricle (LV), right ventricle (RV) and myocardium (MC) are used for cardiac function analysis. Advances in deep learning networks have produced state-of-the-art results. These networks are trained on the FS images and, testing the network with ZF images will result in an unsatisfactory segmentation. We note that a better reconstruction, which is close to the FS image would result in better segmentation performance. In this experiment, we would like to show that the segmentation performance on the reconstructed images from our Recon-GLGAN model is better than the baseline GAN model. 
\begin{figure}
    \centering
    \includegraphics[width=0.8\linewidth]{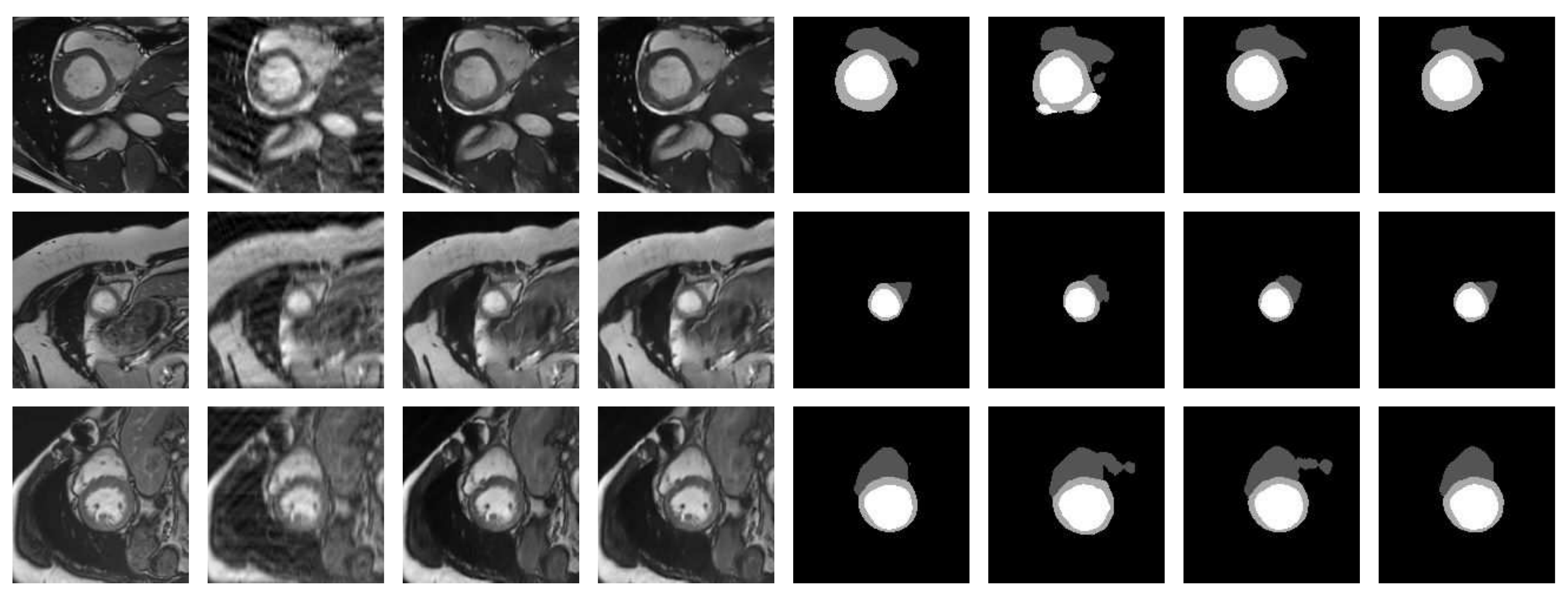}
    \caption{From Left to Right: FS image, ZF image, GAN reconstructed image, Recon-GLGAN reconstructed image, Ground Truth FS segmentation mask, Segmentation mask for ZF, Segmentation mask for GAN reconstructed image and Segmentation mask for Recon-GLGAN reconstructed image. From top to bottom: Sample 1, 2 and 3}
    \label{fig:cmp seg cGAN and GL-cGAN qualitative}
\end{figure}
To demonstrate this, we use the most widely used segmentation network U-Net \cite{unet}. U-Net is trained on the FS images to produce multi-class (LV, RV and MC) segmentation outputs. Since the ground truth segmentation masks are unavailable for the test set of the ACDC dataset, we instead use the outputs of the FS images in the test set as ground truth. The reconstructed images from GAN and Recon-GLGAN are passed to the UNet and the corresponding segmentation masks are obtained. The obtained segmentation masks for sample images are shown in Figure \ref{fig:cmp seg cGAN and GL-cGAN qualitative}. It is evident from the figure that our network's performance is closest to FS followed by GAN and ZF images. The same are quantified using the segmentation metrics Dice and Hausdorff for the sample images in Figure \ref{fig:cmp seg cGAN and GL-cGAN quatitative}. 

\vspace{-5mm}
\section{Conclusion}
% In this paper, considering the application perspective of MRI, we proposed a well thought out gan based architecture GL-cGAN with generator and context discriminator which considers both global and local context. We have conducted experiments to show that our network GL-cGAN produces better reconstruction results compared to baseline methods. We also experimented the possibility of replacing discriminator in gan based reconstruction architectures with our network's context discriminator and found replacement results in improved performance. We also noted that the segmentation of reconstructed images obtained from our network is close to the segmentation of fully sampled images compared to baseline methods.
In this work. we proposed a novel GAN network, Recon-GLGAN. The context discriminator proposed in Recon-GLGAN helps to capture both global and local features enabling a better overall reconstruction. We showed the extensibility of our discriminator with various GAN based reconstruction networks. We also demonstrated that the images obtained from our method gave segmentation results close to fully sampled images. 
% In this work, we proposed a novel GAN based architecture, Recon-GLGAN for MRI reconstruction. Recon-GLGAN consists of a generator and a context discriminator, which  incorporates global and local contextual information from images. Our model offers significant performance improvement over the baseline models. Our experiments show that the concept of a context discriminator can be extended to existing GAN based models to offer better performance. We also demonstrate that the reconstructions from the proposed method also result in better segmentation results.
\newpage
\bibliographystyle{splncs04}
\bibliography{reference}
\end{document}